\pdfoutput=1
\documentclass[
reprint,
amsmath,amssymb,
aps,
pra, 
]{revtex4-1}

\usepackage[normalem]{ulem}
\usepackage{float}
\usepackage{graphicx}
\usepackage{dcolumn}
\usepackage{bm}
\usepackage{xcolor}

\usepackage{physics}
\usepackage{verbatim}

\usepackage{siunitx}
\begin{document}

\title{Critical Entanglement for the Half-Filled Extended Hubbard Model}

\author{Jon Spalding$^{1}$}
 \email{jspal002@ucr.edu}

\author{Shan-Wen Tsai$^{1}$}%
\email{shan-wen.tsai@ucr.edu}

\author{David K. Campbell$^{2}$}%
\email{dkcampbe@bu.edu}

\affiliation{$^1$Department of Physics and Astronomy, University of California, Riverside, California 92521, USA}
\affiliation{$^2$Department of Physics, Boston University, 590 Commonwealth Ave., Boston, Massachusetts 02215, USA}

\date{\today}

\begin{abstract}
We study the ground state of the one-dimensional extended Hubbard model at half-filling using the entanglement entropy calculated by Density Matrix Renormalization Group (DMRG) techniques. We apply a novel curve fitting and scaling method to accurately identify a $2^{nd}$ order critical point as well as a Berezinskii-Kosterlitz-Thouless (BKT) critical point. Using open boundary conditions and medium-sized lattices with very small truncation errors, we are able to achieve similar accuracy to previous authors. We also report observations of finite-size and boundary effects that can be remedied with careful pinning.
\end{abstract}

\pacs{Valid PACS appear here}
\maketitle

\section{\label{Introduction}Introduction}

The one-dimensional Hubbard model is the minimal model for the study of interacting fermions with spin  \cite{oneDhubbard} and has applications in a number of effectively one-dimensional materials including organic conductors, conjugated polymers, and carbon nanotubes  \cite{Ejima2007,ishiguro2012organic,baeriswyl2012conjugated,ishii2003direct} as well as quantum simulators including fermionic cold-atoms \cite{jaksch2005,Baier201,murmann2015two,Gross995} and now quantum dot arrays \cite{hensgens2017quantum}.

 By adding to this model a term for interactions between electrons on neighboring sites, the Hubbard model becomes the Extended Hubbard Model (EHM), which has been simulated using gated quantum dot arrays \cite{hensgens2017quantum}.
 The nearest-neighbor interaction may also be simulated using cold dipolar atoms \cite{2005PhRvL..94p0401G,2012arXiv1204.1725A,2014PhRvL.112a0404A,2011PhRvL.107s0401L,2012PhRvL.108u5301L,2014arXiv1411.3069T} and polar molecules \cite{2008Sci...322..231N,2012PhRvL.108h0405C,Barry2014,McCarron2017,Hummon2013,Anderegg2018,Hemmerling2016} 
 in one-dimensional optical lattices.
 The EHM is described by the Hamiltonian 

\begin{equation}{\label{EHMHamiltonian}}
\begin{aligned}
H_{EHM} ={} & -t\sum_{i,s}(c^{\dag}_{i,s}c_{i+1,s} + c^{\dag}_{i+1,s}c_{i,s}) \\ & + U\sum_{i}n_{i,\uparrow}n_{i,\downarrow} \\ & +
V\sum_{i}n_{i}n_{i+1}
\end{aligned}
\end{equation} 

\noindent
where in second-quantized notation, $n$ represents the site occupancy and $c^{\dag}$ ($c$)  represents a creation (annihilation) operator. This model hosts highly nontrivial many-body physics, even in one dimension, and cannot be studied using analytical means at intermediate coupling. 

\begin{figure}
\centering
\includegraphics[width=0.9\linewidth]{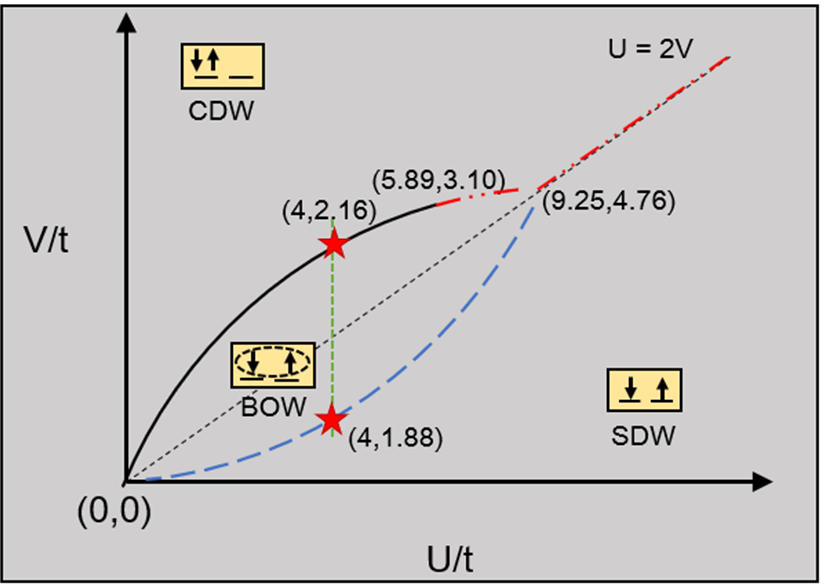}
\caption{[color online] A schematic of the known phase diagram in the repulsive region of the 1D extended Hubbard model. Our study focuses on the two starred critical points, a BKT point at (4,1.88) and a Gaussian transition at (4,2.16). The blue (single dash) line is a BKT transition that spans from the origin to the multicritical point (9.25,4.76). The red (dash-dot-dot) lines represent $1^{st}$-order transitions, and the black (solid) curve is a set of second-order transitions. The black (solid) and red (dash-dot-dot) curves meet at (5.89, 3.10). These values were taken from reference \onlinecite{Ejima2007} for illustrative purposes.}
\label{fig:phasediagramcartoon}
\end{figure}

The phase diagram for the half-filled, repulsive case shown
in figure \ref{fig:phasediagramcartoon}  has been studied and repeatedly updated over four-decades of investigations and became hotly debated once compelling evidence for a thin Bond Order Wave (BOW) region was demonstrated with exact diagonalization \cite{PhysRevB.61.16377} (magnified here for clarity in figure \ref{fig:phasediagramcartoon}). The BOW phase is characterized by a ground state with gapped excitations and alternating bonds between neighboring sites and is separated from a Spin Density Wave (SDW) region by a Berezinskii-Kosterlitz-Thouless (BKT) transition and from a Charge Density Wave (CDW) region by a second-order transition curve that changes at a tricritical point into a $1^{st}$-order transition before terminating at a multicritical point \cite{PhysRevB.93.235118,PhysRevB.61.16377,PhysRevLett.92.236401,zhang2004dimerization,tam2006functional,Ejima2007,glocke2007half,ejima2008ground,menard2011,sengupta2002bond}. In this study, we restrict ourselves to U = 4 in an effort to identify the second-order critical point, herein referred to as $V_{Gauss}$, and the BKT-critical point, $V_{BKT}$ (denoted by star symbols in figure \ref{fig:phasediagramcartoon}).

The phase diagram has been studied with a wide range of methods and has motivated innovations such as parallel tempering for Quantum Monte Carlo (QMC) \cite{PhysRevLett.92.236401}. The studies based on the well-established variational method, Density Matrix Renormalization Group (DMRG) have produced ever-improving results as new measurements have been performed and computations have improved. In 2002, an early DMRG study concluded that the BOW phase appears infinitesimally close to the line U = 2V. This work used the relatively high bond dimension (M) of 1200 and system sizes up to 1024 sites \cite{Jeckelmann2002}. In another DMRG study in 2004, the BKT transition was predicted \cite{zhang2004dimerization} to be at V = 2.01 as extrapolated from moderate (96 to 256) system sizes using a peak in the BOW structure factor, but with the relatively low M of only 500. In 2007, large system sizes (1000) and large M (3000) were used to locate this transition at $V_{BKT}$ $\approx$ 1.877 using standard order-parameter approaches \cite{Ejima2007} which agreed closely with the high-accuracy QMC result of $V_{BKT}$ = 1.89(1) \cite{PhysRevLett.92.236401}\footnote{Our results support $V_{BKT} \approx$ 1.95}. More recently, in 2015, with high M values ($\leq$ 1024) and moderate system sizes ($\leq$ 180) with open boundaries, a careful study used a finite-size corrected spin-gap at U = 4 to get $V_{BKT}$ = 2.08\cite{dalmonte2015gap} which adds controversy to this difficult-to-locate BKT critical point.

A recent study \cite{PhysRevB.96.125129} using a continuous unitary transformation (CUT) approach \cite{krull2012enhanced}
agrees with the numerical values for the CDW/BOW transition and interprets that transition as the condensation of singlet excitons \cite{PhysRevB.96.125129}. 

The phase transitions shown in figure \ref{fig:phasediagramcartoon} have been studied using transition measures based on quantum mechanical many-body properties. Energy-level-crossing methods such as ``fidelity susceptibility'' and ``excited state fidelity'' can accurately identify phase transitions \cite{chen2007fidelity}, and entanglement has been demonstrated as a central tool in the study of many-body quantum physics \cite{amico2008entanglement}. Peaks and discontinuities in various entanglement entropies are useful for models with no a-priori order parameter. The half-chain von-Neumann entanglement entropy (from now on, we refer to the von-Neumann entanglement entropy as simply the ``entropy"), 2-site entropy, and 1-site entropy were previously computed using DMRG to produce an Extended Hubbard model ground state phase diagram \cite{mund2009quantum}. The different methods agreed with Refs.~\onlinecite{Ejima2007,PhysRevLett.92.236401} with some small discrepancies. These discrepancies can, we conclude, be overcome in the EHM using universal results from conformal field theory, previously applied to identification of BKT transitions in the $J_1$-$J_2$ model from the ground state entanglement with periodic boundary conditions (PBC) \cite{Nishimoto2011}. In this paper we extend the method demonstrated in Ref.~\onlinecite{Nishimoto2011} to open boundary conditions (OBC) for the EHM by taking a logarithmic derivative of the entropy for even and odd sites seperately before averaging them to overcome the bond-alternation effects. With this method of computing it, we successfully identify a BKT transition in the EHM with OBC from the peak in the central charge.

Recently, a direct curve fit of the CFT predictions was used to study small lattices, to demonstrate the feasibility of detecting the central charge and the Luttinger exponent directly from the $2^{nd}$ Renyi entropy in cold-atom experiments \cite{kaufman2016quantum}. In Refs.~\onlinecite{PhysRevD.96.034514,PhysRevA.96.023603}, CFT predictions were verified for a one-dimensional bosonic Hamiltonian that acts as a quantum simulator for the O(2) model in 1+1 dimensions, using the midpoint of the chain as  the optimal location to sample the open-boundary DMRG ground state 
because there the finite-size effects as well as boundary effects are minimized, a feature previously exploited in Ref.~\onlinecite{tsai2000density}. However, extracting useful information at the chain midpoint requires a large number of system sizes.

Likewise, it may be prohibitive to repeat an experiment with multiple system sizes, and one-dimensional lattice experiments will usually have a symmetric but inhomogeneous confining potential. So for any numerical or experimental 1D critical models with open boundaries, especially with symmetric but non-uniform potentials, the methods we develop below, which we call ``scaling to the middle,'' should be of value for extracting the most accurate measurements at the midpoint. In short, we re-fit the universal CFT formula for entropy at a 1D quantum critical point to open boundary entropy data for every possible domain centered on the chain midpoint, before extrapolating the curve fit parameters to a domain of 0. This is effectively scaling the curve fitted values in the size of the system block. For the EHM, we combine this curve-fitting algorithm with a simple variance minimum for the CFT curve fit to identify a Gaussian critical point ($V_{Gauss}$) with high accuracy for small system sizes. Compare our value of $V_{Gauss}$ = 2.158 (2.160) from a 64 (128)-site lattice OBC calculation to the best published values of 2.160 from 1000-site QMC \cite{PhysRevLett.92.236401} and 2.164 from 1000-site DMRG \cite{PhysRevLett.99.216403}. We postpone further application and validation of the method, including inhomogeneous potentials, to a future work focused on a simpler model.

In this study, we demonstrate our approaches to finding critical points with OBC ground states and apply them 
to the EHM at half-filling with a cut along the phase diagram at U = 4. Along the way, we expand upon the method developed in Ref.~\onlinecite{Nishimoto2011} for identifying BKT critical points, but for open-boundary wavefunctions, demonstrated by identifying $V_{BKT}$ for our model. Lastly, we characterize the nature of finite-size and boundary effects that occur for this model at $V_{Gauss}$ and in the CDW phase. This includes observations of a degeneracy-induced charge soliton that increases the CFT central charge from 1 to 2 at $V_{Gauss}$, and simple on-site U pinning to eliminate it for both OBC and PBC. We also observe a growth of entropy oscillations away from open boundaries at $V_{BKT}$, contradicting the usual decay of oscillations as observed for Luttinger Liquids.

\section{Methods}

The existence of a mapping between classical critical points in two dimensions and quantum critical points in one dimension implies that the results of conformal field theory also apply for one-dimensional quantum critical points \cite{ElPhCritPhen,QuantPhTran,ScalingRenorminStatMech}.

Using this mapping and field theory techniques, it was shown \cite{JStats2004P06002} that the entanglement entropy of quantum critical points follows, and for open boundaries, the ground state entanglement entropy is \cite{JStats2004P06002}

\begin{equation}\label{SemiInfiniteEntropy}
S_{vN} = S_0 + \frac{c}{6}\log{(\frac{2L}{\pi}\sin{\frac{\pi x}{L}})}
\end{equation}

For periodic boundaries, the factor of 1/6 is replaced with a factor of 1/3. It was later shown numerically that the entropy takes the form  \cite{PhysRevLett.96.100603}

\begin{equation}\label{OpenBoundariesFinite}
S_{vN} = S_0 + \frac{c}{6}\log{(\frac{2L}{\pi}\sin{\frac{\pi x}{L}})} + \frac{\alpha (-1)^x }{(\frac{2 L}{\pi}\sin{\frac{\pi x}{L}})^{K/2}}
\end{equation}
\noindent
for small systems with open boundaries~\footnote{This is difficult to study analytically}. In this updated equation, not only do periodic boundaries change the 1/6 to a 1/3, but also the 2L to L and the K/2 to K \cite{PhysRevD.96.034514}. The coefficient $\alpha$ is non-universal. These details are important for interpreting numerical results, and there are further modifications for generalized Renyi entropies, although the overall form remains the same.
Note that the third term predicts a decay of oscillations away from the boundary, with a universal exponent K called the Luttinger exponent. The Luttinger exponent appears analytically in the weak-coupling bosonization treatment of equation (\ref{EHMHamiltonian}) \cite{tam2006functional}. Even though the analytical bosonization treatment fails at intermediate couplings, the Luttinger Liquid picture is expected to hold in all the critical phases we studied.

\subsection{ The \textit{Scaling to the Middle} method for optimized measurements}

Since the DMRG is best with open boundaries, but
open boundaries induce various edge effects, it is desirable to take measurements at or near the midpoint of a lattice \cite{PhysRevA.96.023603,PhysRevD.96.034514,tsai2000density}.  Many open-boundary effects may be improved by performing measurements at the midpoint for many N and then scaling in N 
\cite{PhysRevLett.96.100603}.

Here we test a complementary approach that improves the accuracy for any single-system-size curve fit measurement 
performed on open boundary condition data \cite{PhysRevLett.104.095701,PhysRevLett.96.100603,FriedelOsc2002}.

\begin{figure}
\centering
\includegraphics[width=0.9\linewidth]{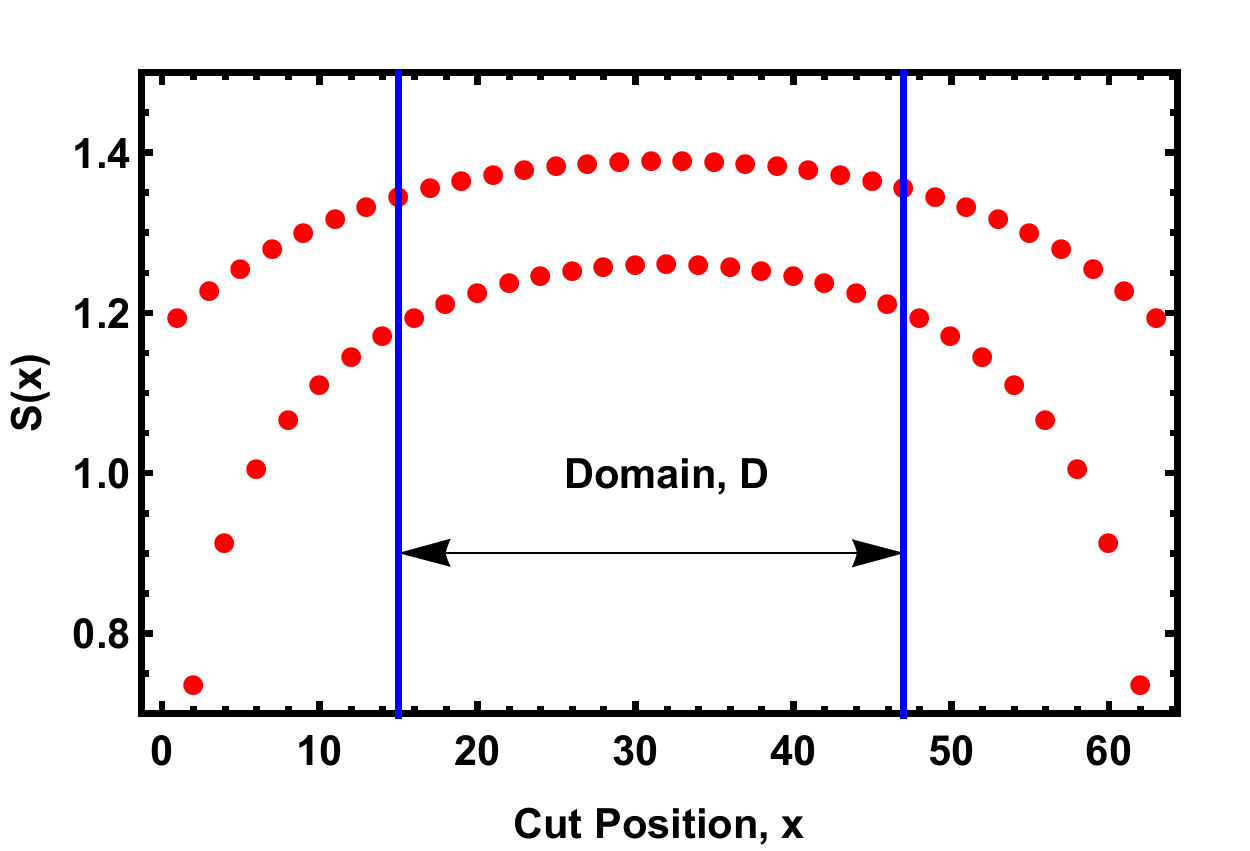}
\caption{[color online] The definition of the ``Domain, D" when curve fitting the entanglement entropy. This entropy was computed from the ground state for U = 4 and V = 1, and compares well to the expectations for a critical spin chain~\cite{PhysRevLett.96.100603}.}
\label{DomainFig}
\end{figure}

\begin{figure}
\centering
\includegraphics[width=0.9\linewidth]{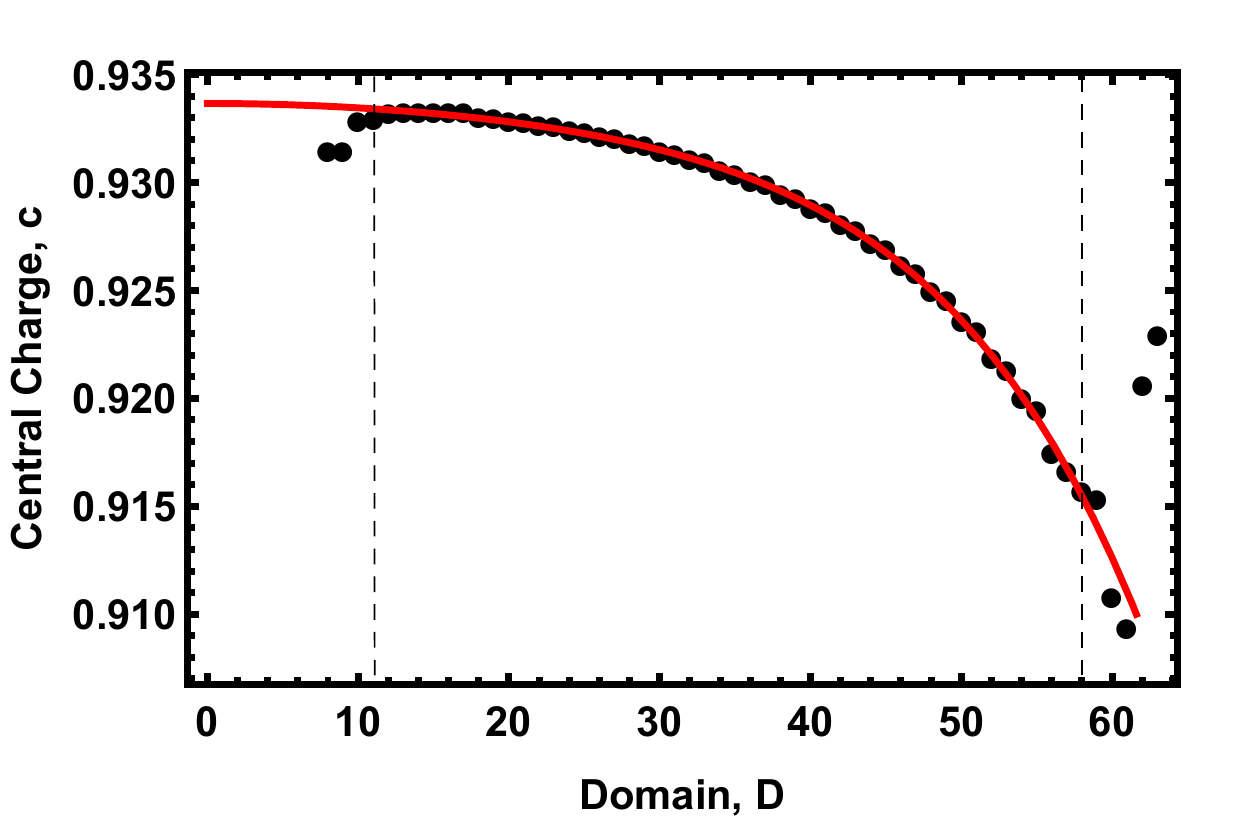}
\caption{The ``Scaling to the Middle'' approach to extracting finite-size measurements from open-boundary condition data illustrated for the central charge when U = 4 and V = 1 for a 64-site entropy dataset. A curve fit is performed for each domain, D as in figure \ref{DomainFig}, which results in a fitted value of central charge, $c$. The values are extrapolated to domain 0 to produce a best estimate value. Here, the $6^{th}$-order polynomial fit to $c(D)$ shows that $c$ = 0.934.}
\label{CvsDfigure}
\end{figure}

We illustrate the method by computing central charge for a 64-site lattice in the critical SDW phase, which agrees well with previous studies of critical spin models \cite{PhysRevLett.96.100603}. Figure \ref{DomainFig} shows the centered domain D, which is curve fitted by equation \ref{OpenBoundariesFinite} to extract a value of c(D). This is repeated for all D before fitting the values of c vs D using an even function. At the end of this procedure, the value extrapolated to D = 0 represents the ``best value'' for this lattice size as illustrated in figure \ref{CvsDfigure}. Note that overfitting and strong edge effects, when D is too small or too large respectively, restrict which values of D are used in the curve fit.

\begin{figure}
	\centering
	\includegraphics[width=0.9\linewidth]{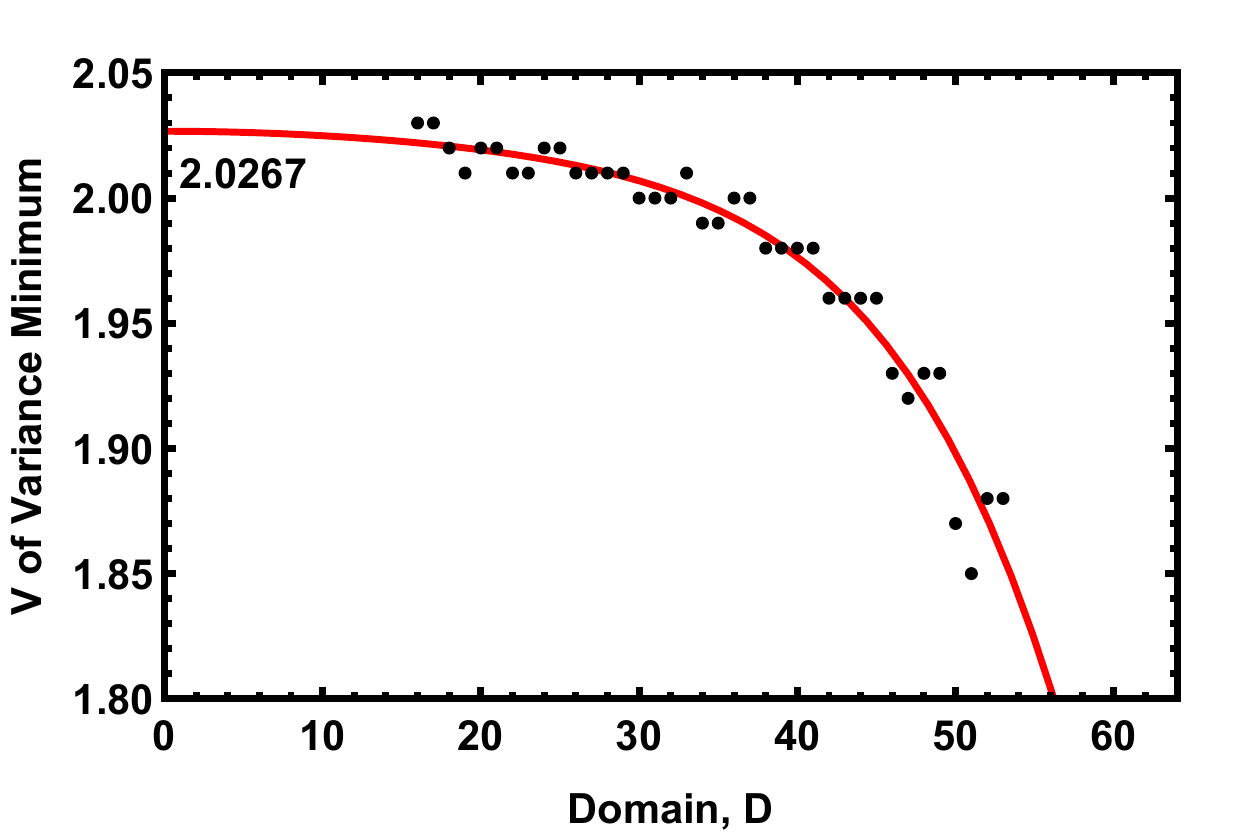}
	\caption{This plot of variance minimum vs domain was produced using data for 64 sites and illustrates that the variance minimum does not work for equation \ref{OpenBoundariesFinite} when applied at the BKT point. We would expect the best estimated $V_{BKT}$ to be about 1.82 instead of about 2.03 as shown in this figure. This also illustrates the utilitiy of ``Scaling to the Middle" for rapidly uncovering finite size and boundary effects.}
	\label{BKTscalemiddle}
\end{figure}

Lastly, we comment that figure \ref{BKTscalemiddle} demonstrates the utility of ``scaling to the middle'' in checking finite-size and curve-fit domain effects of a given measurement, and is consistent with the physicists' standard tool of scaling in system size.

\subsection{The \textit{Variance minimum} method for finding critical points}
\begin{figure}
	\centering
	\includegraphics[width=1\linewidth]{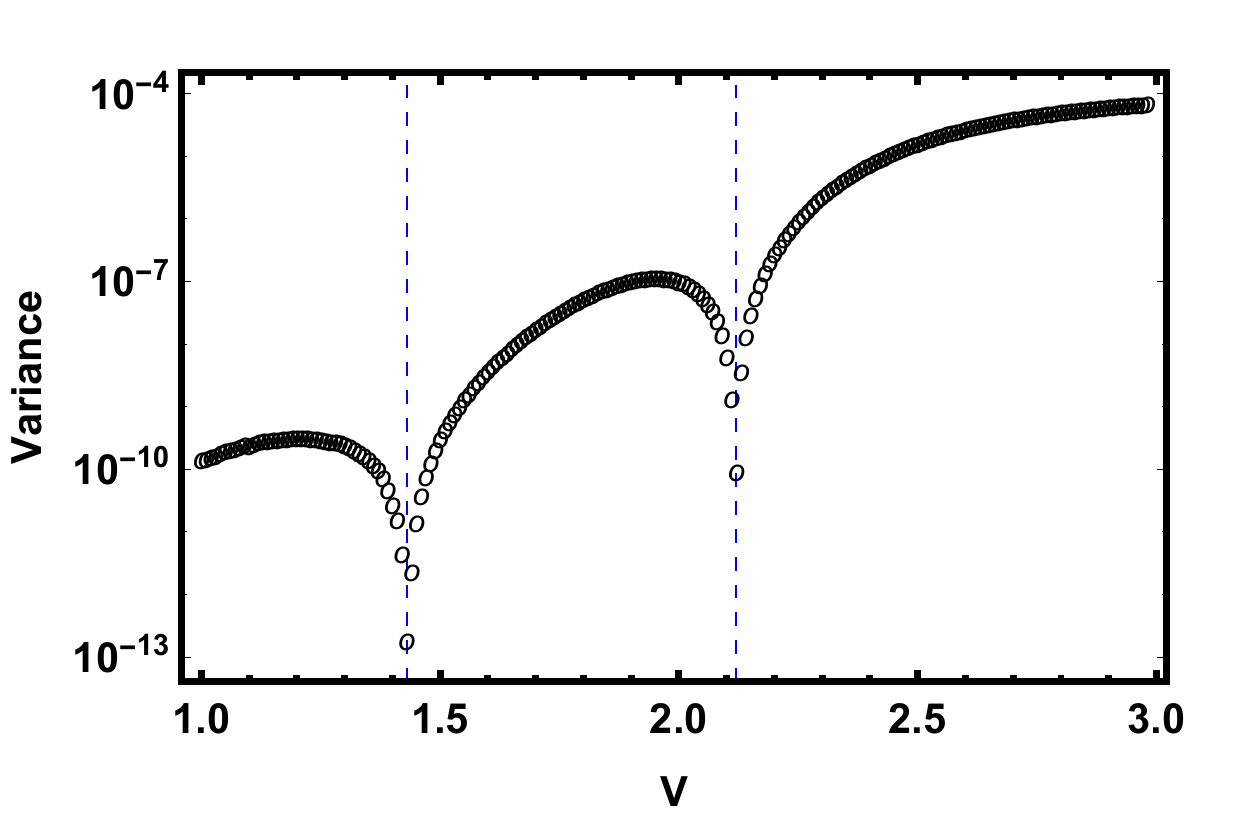}
	\caption{Curve fit variance vs V for 16 sites, with two minima at 1.43 and 2.12. 2.12 is 0.04 above the result published in Ref.~\onlinecite{zhang2004dimerization} which were generated using much larger system sizes. Note that a single domain, D = 8, was used to produce this plot.}
	\label{naivefigure}
\end{figure}

\begin{figure}
\centering
\includegraphics[width=0.9\linewidth]{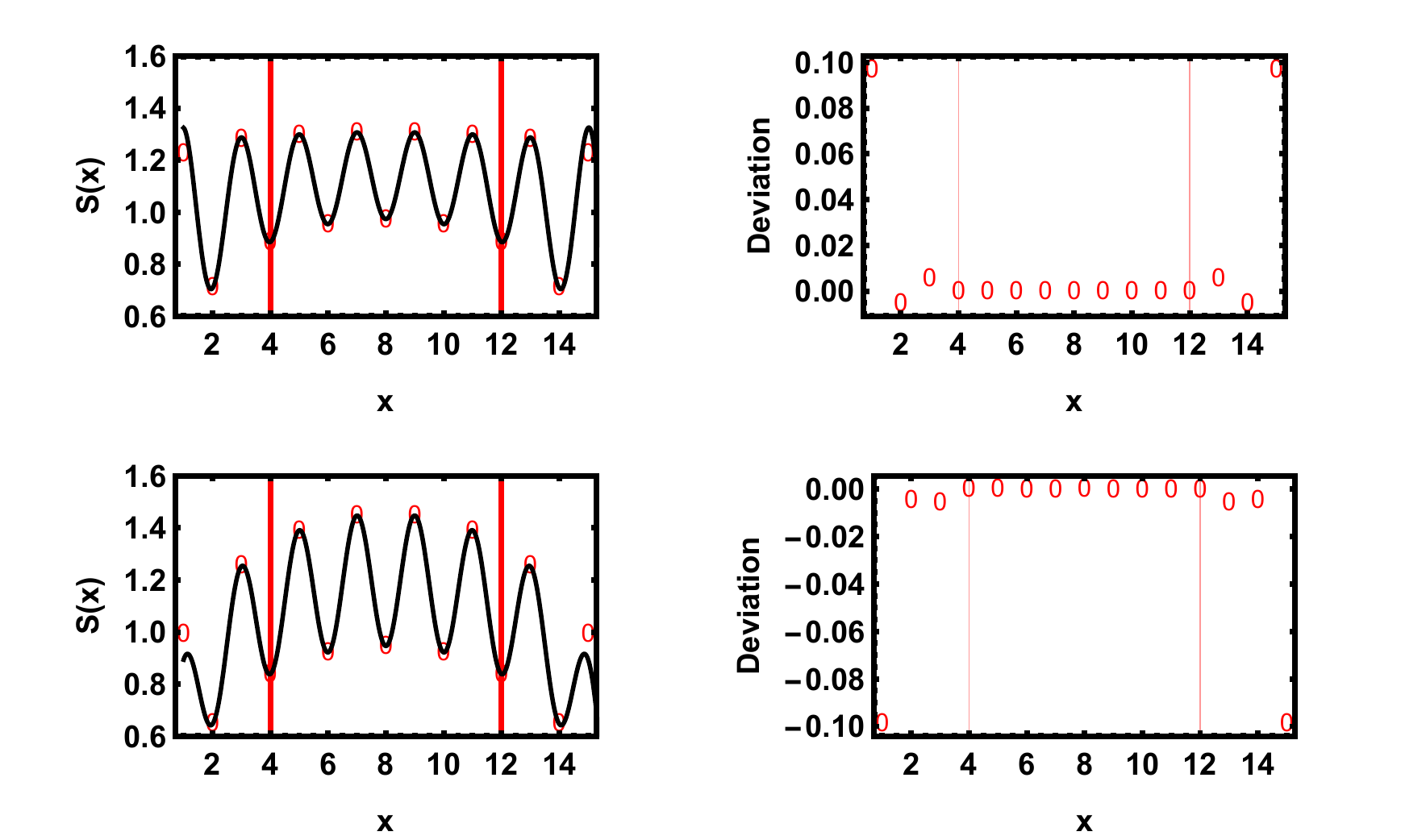}
\caption{The entropy for a 16-site lattice plotted along with the CFT curve fit at the two variance minima in figure \ref{naivefigure}, V = 1.43 (top) and 2.12 (bottom), respectively. At right are plots of the deviations of the curve fit from the data. Note that the curve fit was restricted to the domain of data between the red vertical bars.}
\label{deviationsplot}
\end{figure}

The conformal entropy formula, equation \ref{OpenBoundariesFinite}, only fits at critical points, which implies that a plot of the variance vs. coupling constants along a cut in the phase diagram will exhibit a clear minimum when such a critical point separates two gapped phases (for instance, along U = 4 from BOW to CDW). This works very well for all of the system sizes we studied and provides an extremely sharp, reliable transition indicator, with very low error even for small system sizes, as illustrated for 16 sites in figure \ref{naivefigure}. The plot in figure \ref{naivefigure} was generated by fixing D to the middle half of the data, as illustrated in figure \ref{deviationsplot}, which also shows the entropy at the two variance minima from figure \ref{naivefigure}.

\begin{figure}
	\centering
	\includegraphics[width=0.9\linewidth]{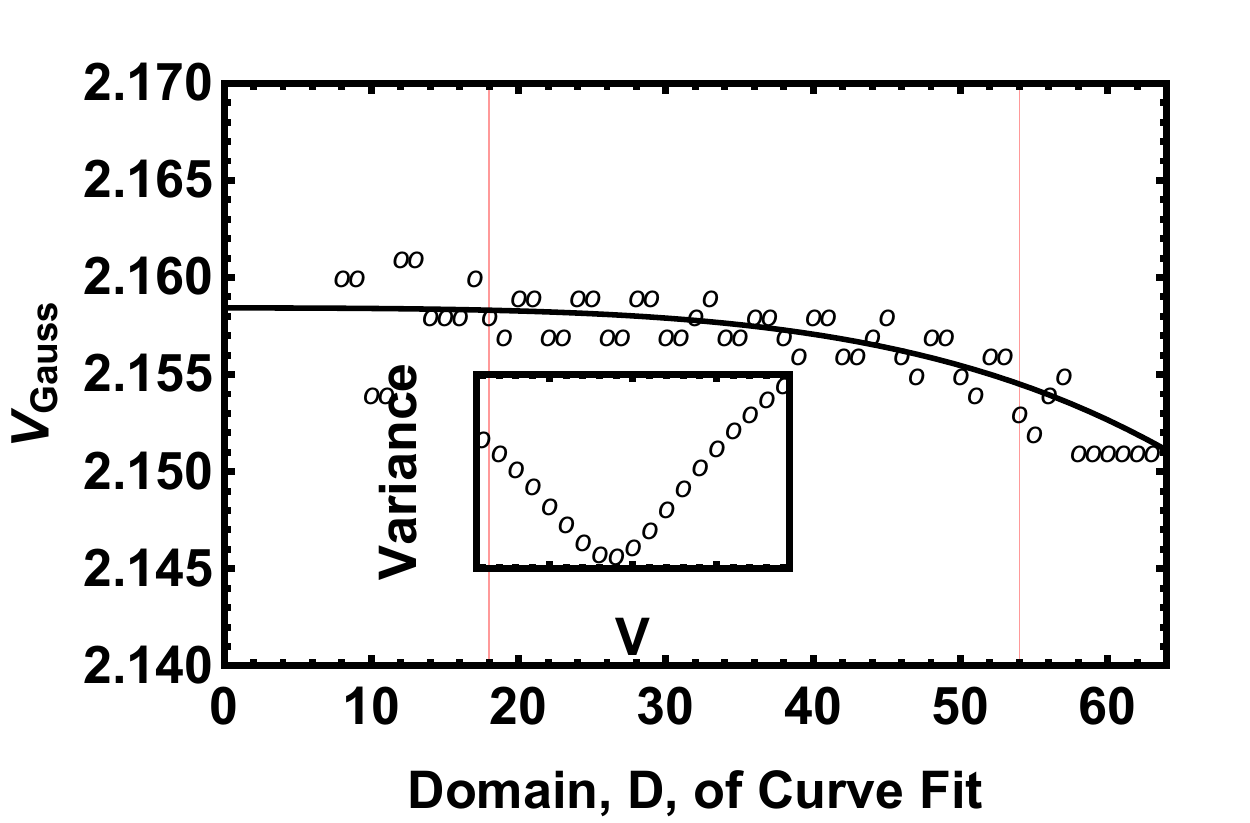}
	\caption{Illustration of the combined ``scaling to the middle" and ``variance minimum''
	procedure applied to identifying a critical point  for 64-site data. The constant term in the polynomial fit is the value of interest; in this case it is the critical point, $V_{Gauss}$. First, for each domain D, the minimum variance is used to identify the critical point (shown in subplot) and these critical points are then fitted as a function of D with an even polynomial. The constant in the curve fit, here 2.158, is the best estimate for the critical point. A conservative error estimate is $\pm 0.001$.}
	\label{upgradefigure}
\end{figure}

We can then combine the ``Scaling to the middle" technique with the ``Variance Minimum" method, as shown in figure \ref{upgradefigure}. Each of the data points in that figure is the $V_{Gauss}$ corresponding to the variance minimum for a particular D (shown in the inset). This collection of critical points is then curve fitted and extrapolated to an effective D of 0. This extrapolation step of the procedure requires care, since one needs to throw out some data. When D is too small, overfitting disrupts the CFT curve fit, and when D is too large, edge effects disrupt the CFT curve fit.

BKT phase transitions divide a critical \textit{region} from a gapped region, so that the variance is not expected to produce a clear minimum, but rather some kind of a step feature. Unexpectedly, we still found 
a minimum in our data (figure \ref{naivefigure}) that we clarified using our ``scaling to the middle" approach for 64 sites in figure \ref{BKTscalemiddle}. Since the value extrapolated by scaling to the middle disagrees with our more reliable results presented below, we conclude that the minimum associated with the BKT transition is not a good transition indicator. Also, for fixed D, scaling the position of this variance minimum in system size shows that it fails to identify the transition. Next, we describe reliable ways of finding $V_{BKT}$ that overcome the failure of the variance minimum approach.

\subsection{Modified log-derivative and c-max for locating BKT transitions with OBC\label{chargemaxsect}}

To compensate for the failure of the variance minimum approach at the BKT transition, we found another approach, proven for ground states with periodic boundaries \cite{Nishimoto2011} which 
we demonstrate, with modifications, for open boundaries. This method \textit{hinges on} the presence of a finite-size correction to central charge, $c$, at BKT points~\cite{JStats2004P06002}$ $\footnote{The method is not well-suited to the study of the thermodynamic limit since the corrections to $c$ decrease as system size (accuracy) increases (decreases)}.

For periodic boundaries, we start from equation \ref{SemiInfiniteEntropy} (with 6 replaced by 3) and take a derivative with respect to the logarithm, evaluated at the middle of the chain. The result is an equation for the central charge:

\begin{equation}\label{PBCCentralCharge}
c(x) = 3 \derivative{S_{vN}(x)}{\log{(\frac{2L}{\pi}\sin{\frac{\pi x}{L}})}}
\end{equation}

\noindent
which simplifies, for x = L/2 on a discretized lattice, to~\cite{Nishimoto2011}

\begin{equation}\label{PBCCentralChargeSimple}
c(L/2) = 3 \frac{S_{vN}(L/2-1) - S_{vN}(L/2)}{\log{\cos{(\frac{\pi}{L}})}}
\end{equation}

This simplified form applies only when there is no oscillatory term, such that the numerical derivative works for nearest neighbor bonds.

Since open boundaries, and higher Renyi index, will both induce oscillations in the entanglement, we propose to use the modified version based on equation \ref{OpenBoundariesFinite}, in which the finite differences are evaluated on next-nearest-neighbor sites (or $n^{th}$-order neighbors for longer, but still commensurate, wavelength oscillations)~\cite{PhysRevB.93.235118}$ $\footnote{This was also used in reference \cite{PhysRevB.93.235118} but for PBC}. The result is

\begin{equation}\label{OBCentralCharge}
c(x) \equiv 6 \frac{S_{vN}(x+1) - S_{vN}(x-1)}{\log{\sin{(\frac{\pi(x+1)}{L}})}-\log{\sin{(\frac{\pi(x-1)}{L}})}}
\end{equation}	

Two complications arise in this approach: first, even-numbered sites produce different values of $c(x)$ than odd sites, and second, equation \ref{OBCentralCharge} can behave poorly near the middle bond of the chain ($x = N/2$) due to inexact canceling of a 0 in the numerator and denominator.

\begin{figure}
	\centering
	\includegraphics[width=0.9\linewidth]{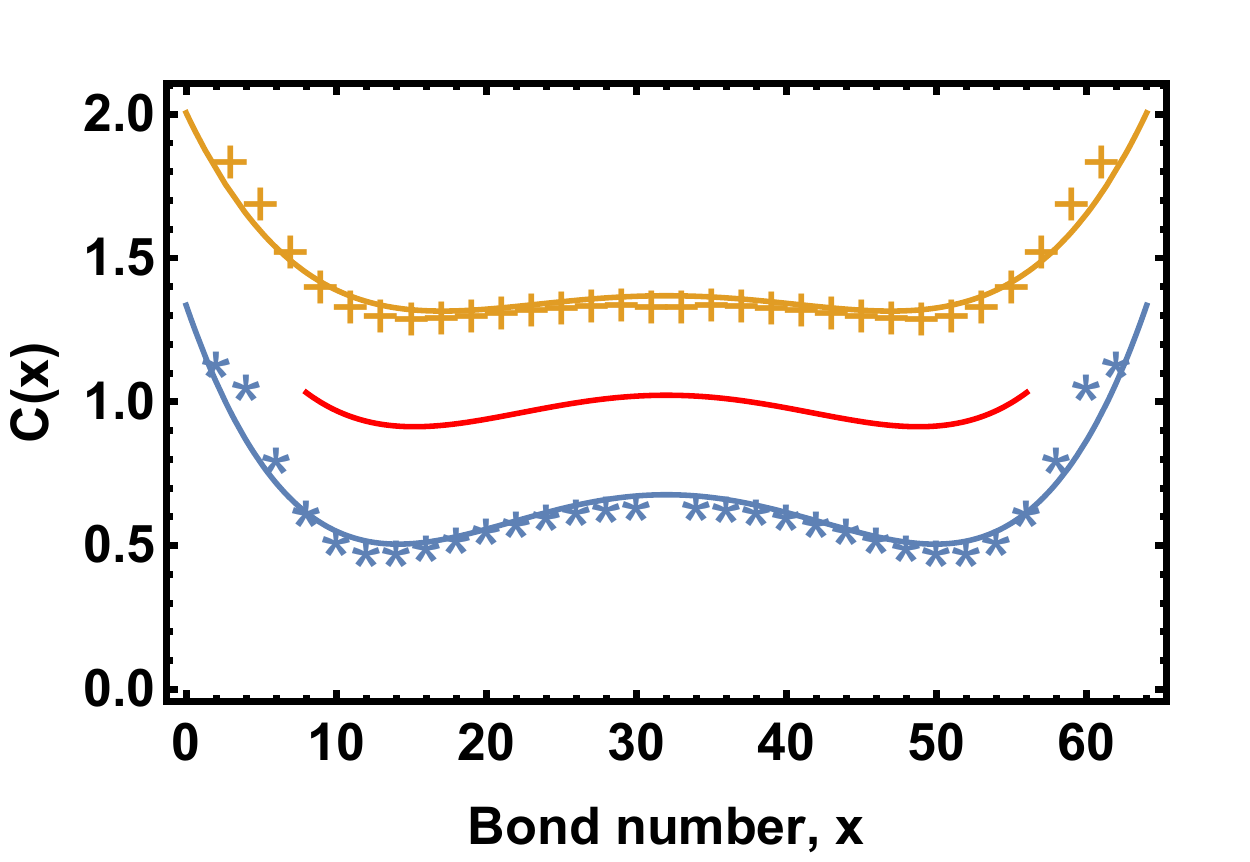}
	\caption{[color online] Illustration of the determination of central charge as a function of position for a 64 site lattice at $V_{BKT}$ = 1.83. The yellow pluses and yellow curve fit correspond to the even bond log-derivatives while the blue stars and curve fit correspond to the odd bond log-derivatives. Note that the midpoint blue star was removed due to an infinity. Lastly, the red (middle) curve is the resulting $c(x)$ averaged from the yellow (upper) and blue (lower) curves. As with the other curve-fitting approaches presented in this paper, the fluctuations at the edges were excluded from the curve fit.}
	\label{c_eff_x}
\end{figure}

We resolve the first difficulty by curve fitting $c_{even}(x)$ and $c_{odd}(x)$ separately, and then averaging the curve fits to produce a single function of x. We resolve the second difficulty by inspecting the data by eye to find aberrant values of $c(x)$ at the chain midpoint that are excluded from the curve fit. In practice, we cut out from 1 to 3 data points for every entropy dataset. The resulting curve, evaluated at $L/2$, provides our best estimate of c for a given system size $L$. This process is illustrated for entropy data in figure \ref{c_eff_x}.

Lastly, as was done in Ref.~\onlinecite{Nishimoto2011}, we used the maximum value of $c(V)$ to indicate the BKT transition. This method appears to work for our model but with large finite-size effects, and the resulting $V_{BKT}$ is consistent with previous work. With refinement of the method for OBC (see future publication), we should be able to greatly improve the precision of $V_{BKT}$ for the EHM.

Although we did not use scaling in domain size (i.e. re-fit $c(x)$ for every possible domain of the data D), varying the domain D did provide an estimate of the error in c and the critical point, as reported in table \ref{EntropyDerivative}.

\section{\label{Results}Results}

In this section, we report our observations for many system sizes, and discuss important finite-size effects including a charge soliton that creates an effective second bosonic degree of freedom at $V_{Gauss}$ for small lattices, and open-boundary charge oscillations that can be removed with tuned boundary softening.

\subsection{Identification of second-order transition}
First we summarize our efforts to identify $V_{Gauss}$ using the combined variance minimum and scaling to the middle method, presented in table \ref{upgradetable}.

To quickly review how table \ref{upgradetable} was produced, for each system size, and each domain, we identified a critical point from the minimum in the variance. Then, for each system size, we used ``scaling to the middle'' to get a best estimate of the critical point at an effective domain size of 0. This procedure is illustrated in figure \ref{upgradefigure}.

One of the advantages of this approach is that it implicitly provides an error estimate for the measurements taken for a given system size. The errors we report in table \ref{upgradetable} are estimated conservatively from the plot of a parameter versus fit domain or, if the plot is particularly smooth, by using the polynomial curve fit error. Depending on the discretization of V, this plot can either jump erratically, as $V_{Gauss}$ changes with changing domain, or it can trend smoothly towards a very clear result. For instance, for 64 sites, the procedure is illustrated in figure \ref{upgradefigure} which shows that the discretization of V, 0.001, is a good estimate of the error in the extrapolated value $V_{Gauss}$ = 2.158.

\begin{table*}
	\caption{CFT curve fit results at the Gaussian critical point as determined by combining the \textit{variance minimum} and \textit{scaling to the middle} Observe that K remains negative as the curve fit domain is scaled to the chain midpoint, and that it becomes \textit{more} negative for larger system sizes. Note that $S_0$ and $A \equiv -\frac{\alpha}{(2L/\pi)^{K/2}}$ are non-universal, and that the oscillatory term is inverted (K is negative) from what is expected for Luttinger Liquids \cite{PhysRevLett.96.100603}. Also note that $\alpha$ is strongly dependent on system size and is a non-universal curve fit parameter. The \textit{numerical} resolution on V was 0.001 for all system sizes in this table. We report the estimated error in the last significant figure in parenthesis. The DMRG precision was limited by the values in \textbf{bold}; for small systems, M was unbounded, while for large systems, M was fixed. $\Delta$E as reported here is a conservative estimate on the accuracy of the ground state energies achieved in our DMRG calculations. At $V_{Gauss}$, the soliton increased entanglement so that only system sizes 16 and 32 were nearing exact diagonalization precision.}
	\begin{ruledtabular}
		\begin{tabular}{rdddddccc}

		\textbf{N}    & \multicolumn{1}{c}{\textbf{$V_{Gauss}$} }     & \multicolumn{1}{c}{\textbf{$S_0$}}   & \multicolumn{1}{c}{\textbf{c}}       & \multicolumn{1}{c}{\textbf{$A$} }   & \multicolumn{1}{c}{\textbf{K}}        & 
		\textbf{M} &
		\textbf{trunc.}  &
		\textbf{$\Delta$E} \\
		\hline
		16   & 2.12~(2)     & 0.42(7)    & 2.0~(2)  & 0.25   & -0.16    & 900  & \textbf{5E-14}  & 3E-12 \\ 
		32   & 2.150~(5)  & 0.30~(3) & 2.12~(5) &  0.18 & -0.5~(1)  & 2000 & \textbf{5E-14}  & 3E-11\\
		64   & 2.158~(1)  & 0.31~(1) & 1.97~(1) & 0.14 & -0.81~(1)   & \textbf{3200}  & 1E-13 &  5E-9\\ 
		128  & 2.1605~(5) & 0.41~(5) & 1.71~(5) &  0.12 & -1.07~(2)   &  \textbf{3200} & 1E-11 &  3E-7 \\
		256  & 2.160~(5) & 0.65~(5) & 1.4~(1) &  0.10 & -1.15~(5)  & \textbf{3200}  & 1E-10 &  3E-6 \\ 
		\end{tabular}
\end{ruledtabular}
	\label{upgradetable}

\end{table*}

Looking more closely at table \ref{upgradetable}, two of the results are quite surprising. First, the central charge starts at the (unexpectedly high)\cite{JStats2010P04023} value of 2 before trending downwards with increasing system size, and second, the Luttinger exponent decreases passed -1, when we expect the result to be 0.44 \cite{PhysRevLett.92.236401}. We next attempt to clarify these observations.

\begin{figure}
\centering
\includegraphics[width=0.9\linewidth]{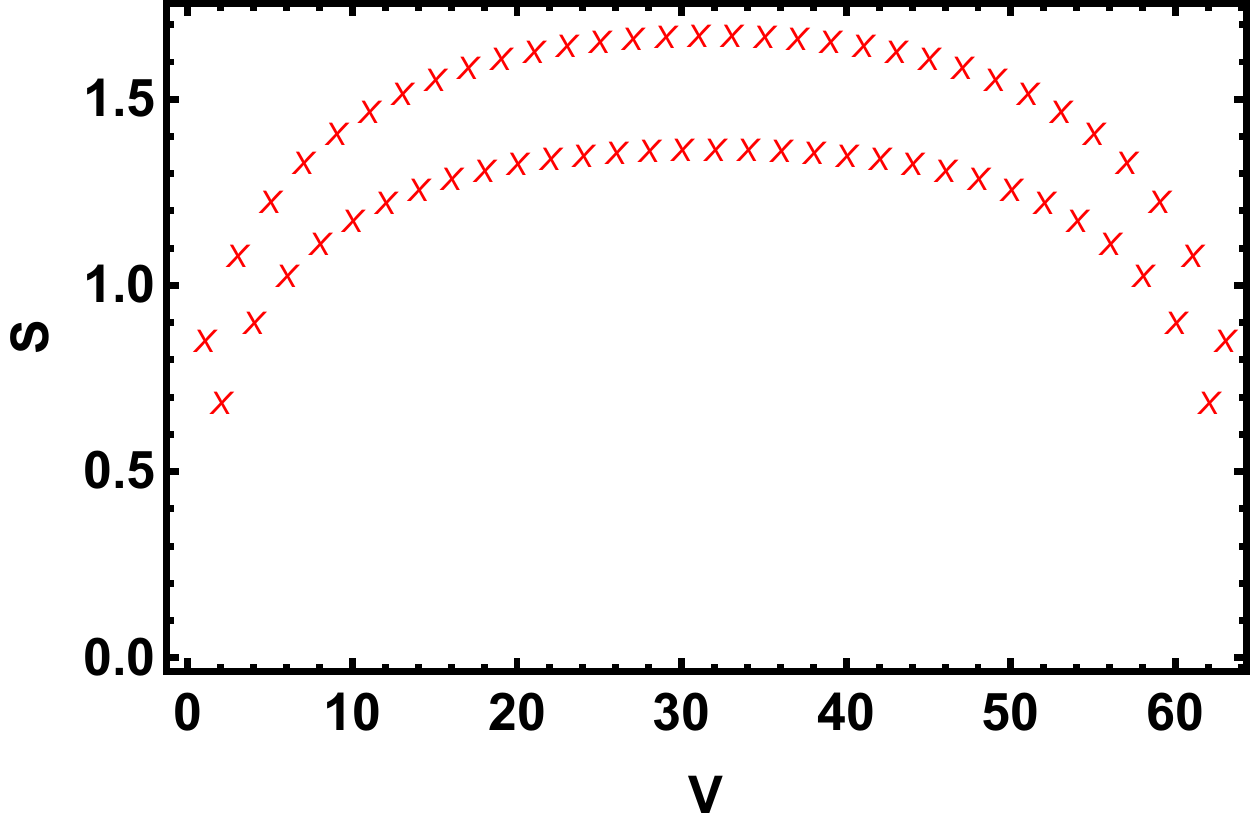}
\caption{Entanglement as a function of bond number for 64 sites at the critical point, $V_{Gauss}$ = 2.158. The entanglement envelope function has unexpected charge effects. Oscillations grow from the boundaries to a maximum in the middle, and the averaged curvature is greater than expected for a c = 1 theory.}
\label{finitefigure}
\end{figure}

\begin{figure}
	\centering
	\includegraphics[width=0.9\linewidth]{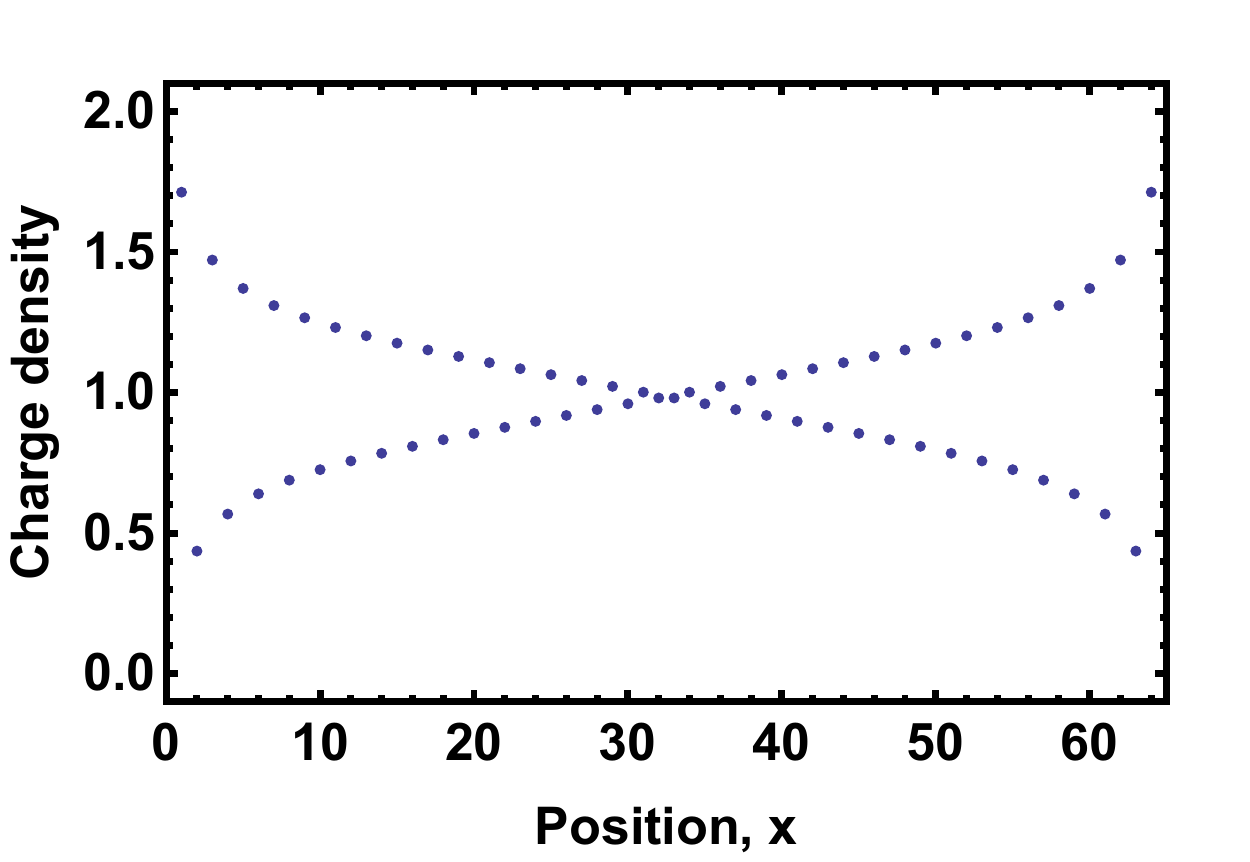}
	\caption{Charge density as a function of position at V = 2.160 for 64 sites. Charge oscillations appear at the boundaries with a long-range decay, distinct from the charge solition in the CDW phase in figure \ref{solitondensity}.}
	\label{ChargeDensity}
\end{figure}

We start by looking directly at a plot of the entanglement entropy as a function of position at $V_{Gauss}$ in figure \ref{finitefigure}. The most obvious feature is that the oscillation growth has been reversed from that expected at a critical point exhibiting Luttinger Liquid criticality, for instance as shown in figure \ref{DomainFig}. Ordinarily, the oscillations in entropy decrease towards the middle of the lattice \cite{PhysRevLett.96.100603}, but in this case they clearly grow. As can be seen in table \ref{upgradetable} this behavior does not change with larger system sizes or scaling to the middle -- it appears to be a feature of this critical point with open boundary conditions. However, the maximum oscillation amplitude, which happens to be at the midpoint of the lattice, \textit{does} decrease with increasing system size, just as the midpoint oscillation amplitude decreases with system size in a Luttinger Liquid. 

To elucidate whether the oscillation inversion or central charge increase behavior is a remnant of the finite-size behavior of the CDW phase, we studied the entropy in that region of the phase diagram (that is, $V > V_{Gauss}$) and compared it with our results at $V_{Gauss}$. 

\begin{figure}
\centering
\includegraphics[width=0.9\linewidth]{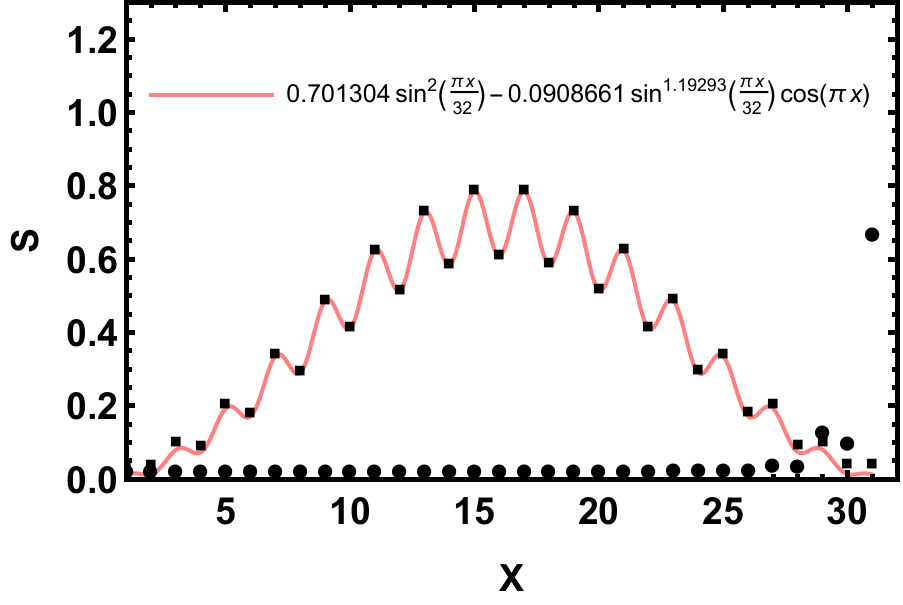}
\caption{The entropy in the CDW phase fits well to a combination of sine functions, as shown here for 32 sites at U = 4 and V = 10. The square points are raw entropy data produced without pinning, while the round points are raw entropy with pinning. Compare the exponent for the second term, 1.2, to the negative of the exponent K in table \ref{upgradetable}, which is approaching 1.2 for large sizes. The pinning for this DMRG calculation was U = U + 1.0 at the left edge and U = U - 1.0 at the right edge.}
\label{solitonentropy}
\end{figure}

\begin{figure}
\centering
\includegraphics[width=0.9\linewidth]{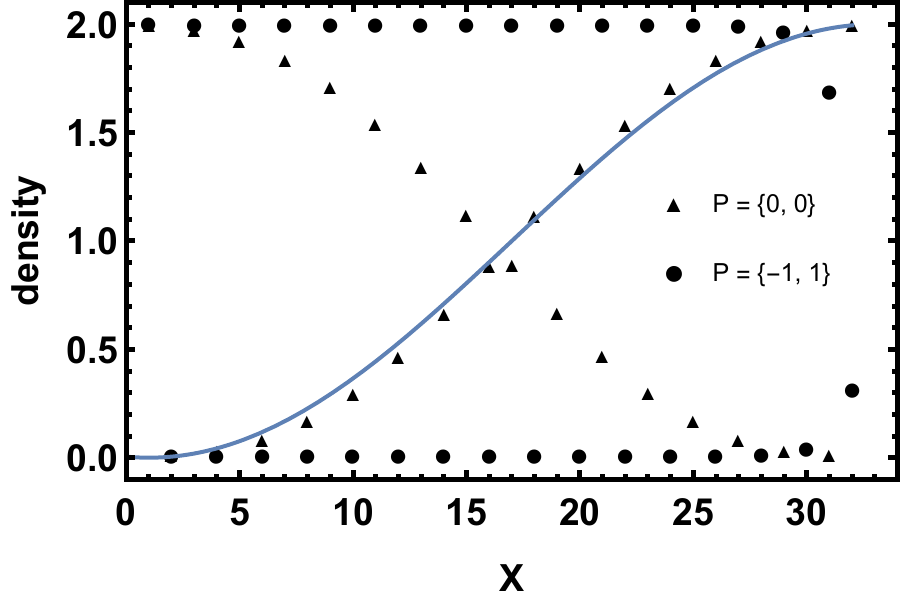}
\caption{Plot of charge density for N = 32 sites, U = 4, V = 10, deep in the CDW phase with ends pinned with U + P where P is 0 or $\pm$ 1. Here we see that pinning eliminates a topological defect (a kink-soliton). The envelope of the oscillating soliton density fits approximately to 2 $\sin{\left(\frac{\pi (x - 1)}{2L}\right)}^2$ as shown.}
\label{solitondensity}
\end{figure}

It turns out that the ground state energy is minimized in the CDW phase when two degenerate phases are present with a $\pi$ phase shift between the two ends for OBC. This causes a topological defect with associated entropy plotted in figure \ref{solitonentropy} and density plotted in figure \ref{solitondensity}. The entropy and density both fit well to combinations of sine functions as shown. Still considering the CDW phase, PBC also has a uniform nonzero entanglement entropy due to the soliton/degeneracy effect.

For OBC, increasing the on-site energy U at site 1, while decreasing it at site L, is effective at picking out one of the degenerate states and eliminating the soliton as shown in figures \ref{solitonentropy} and \ref{solitondensity}. Likewise, for PBC, the nonzero entropy is lowered to 0 (i.e. a classical CDW) by increasing or decreasing U at a single site.

Now considering $V_{Gauss}$, the oscillatory inversion is caused by charge-density oscillations as seen in the plot of density, figure \ref{ChargeDensity}; with fine-tuned pinning, in this case with additional \textit{positive U at both site 1 and site L}, we were able to recover both a central charge close to 1 (best result was 1.2 for 64 sites) and regular Luttinger Liquid effects (K was about 0.5 for 64 sites, close to the previous Monte Carlo best estimate of 0.44). We pursued pinning to check that we could separate the boundary-induced effects from the critical behavior. Fine-tuned pinning softens the boundary conditions and reduces the oscillations, hinting at further work to be explored with alternative boundary conditions \cite{PhysRevLett.71.4283} which may combine nicely with ``scaling to the middle'' to extract accurate infinite-size values.

The growth of the entropy oscillations we present in figure \ref{solitonentropy} and table \ref{upgradetable} was also displayed in Ref.~\onlinecite{PhysRevA.95.053614} in a different model with charge oscillations, however the authors did not investigate the growth of the oscillations from the open boundaries.

We also studied a 16-site lattice at $V_{Gauss}$ with PBC. Oddly, although the oscillations no longer existed, we still found an increased central charge of 2, which further supports the presence of a soliton as a finite-size effect at this critical point. In the CDW phase ($V > V_{Gauss}$), for periodic boundaries, the soliton was eliminated easily by increasing U at a single site. This strategy worked at $V_{Gauss}$ as well, bringing the central charge down to the thermodynamic-limit value of 1, while inducing a small charge oscillation. We hypothesize that the soliton is contributing a second bosonic degree of freedom for small systems, an effect that should be observable experimentally.

For open boundary conditions, we can see from our data in table \ref{upgradetable} that the central charge of 2, and hence the soliton, is largely unchanged until the system size reaches about 100 sites. This is encapsulated by the approximate scaling of c with N, according to the function $c(N) \approx 1 + \tanh(100/N)$. To arrive at this function, we included preliminary calculations of large (512 and 1024) site systems. These calculations also showed the oscillation growth from the boundaries. However, our data was incomplete and had low convergence relative to our other data, so we chose to hold back on reporting these results, as promising as they were \footnote{In order to accurately study the critical behavior, we need to first find the critical point precisely. Since the critical point sharpens as system size increases, and since computational cost grows rapidly with system size, the necessary scan across many V values is very computationally demanding.}.

\subsection{\label{BKT}Identification of the BKT Transition}

\begin{table*}
\caption{Critical point and resulting curve fit parameters as determined by fitting the entanglement entropy, equation \ref{OpenBoundariesFinite}, and applying the ``scaling to the middle'' approach for all parameters. The maximum in $c(V)$ was used to identify the critical point. $A$ was defined above in table \ref{upgradetable}. Note that for 16 sites, scaling to the middle is difficult due to the small system size (hence the values of Not Available (NA)).  Also, for sizes 16 and 32, we did not record the maximum bond dimensions. The bolded values were used to set the DMRG convergence. When the truncation error (trunc) was used, M was allowed to grow unbounded; when M was fixed (due to resource limitations) sweeps were continued until $\Delta$E or trunc was achieved. For large systems, M is always the limiting factor. At $V_{BKT}$, sizes 16, 32, and 64 are nearing exact diagonalization precision.}
	\centering
	\begin{ruledtabular}
		\begin{tabular}{cdddddccc}
	
		\textbf{N} & \multicolumn{1}{c}{\textbf{$V_{BKT}$}} & \multicolumn{1}{c}{\textbf{$S_0$}} & \multicolumn{1}{c}{\textbf{c}} & \multicolumn{1}{c}{\textbf{$A$}} & \multicolumn{1}{c}{\textbf{K}} & \textbf{M} & \textbf{trunc} & \textbf{$\Delta$E}\\ \hline
		16         &       \multicolumn{1}{c}{NA}             &   \multicolumn{1}{c}{NA}           &     \multicolumn{1}{c}{NA}       &           \multicolumn{1}{c}{NA}     &       \multicolumn{1}{c}{NA}    & NA & \textbf{1E-13} & 1E-12 \\ 
		32         &       1.56~(1)            &         0.776~(5)       &      0.974~(5)      &       0.12~(1)         &      1.193~(5)    & NA & \textbf{1E-13}  &  1E-12\\
		64         &         1.82~(1)           &       0.76~(1)         &     1.0542~(2)       &      0.10~(1)         &     1.125~(5)     & 2000 & \textbf{5E-14} & 5E-10\\
		128        &      1.95~(1)              &       0.780~(5)         &     1.060~(1)       &         0.08~(1)       &       0.98~(1)   & \textbf{3200} & 1E-12 & 2E-8 \\
		256        &           1.93~(1)         &       0.797~(1)         &     1.028~(1)       &      0.06~(1)          &    0.97~(1)    & \textbf{3200} &  1E-11  & 1E-7\\ 
		\end{tabular}
	\end{ruledtabular}
	\label{ScaleToMiddleBKT}
\end{table*}

It has been known for some time that BKT transitions are difficult to detect numerically due to the slow closing of the gap for standard order-parameter and energy gap methods \cite{PhysRevLett.92.236401}. Previous entanglement entropy studies of the EHM's BKT transition have been imprecise: using the two-site and block entropies lead to a discrepancy in $V_{BKT}$ of about 0.1 from the best published results, even though the system sizes were large (512 sites) and the truncation error low (equivalently, high bond dimension M = 3000) \cite{mund2009quantum}. We identified an approach that provides a sharper, more accurate transition indicator, based on the universal scaling law \ref{OpenBoundariesFinite} for the ground-state entanglement.

As demonstrated in Ref.~\onlinecite{Nishimoto2011} and citations to that article \cite{PhysRevB.95.085102,PhysRevB.94.235155,PhysRevLett.113.020401,1742-6596-592-1-012134}, the peak in the central charge provides a reliable, universal way of identifying BKT transitions from finite-size data. We demonstrate this approach for the EHM with two methods: extract central charge for each V with a simple curve fit that has been scaled to the middle, and the logarithmic derivative method to extract central charge for each V, as described in the Methods section, \ref{chargemaxsect}. The results presented below are to be compared against the most reliable, found in Refs.~\onlinecite{Ejima2007,PhysRevLett.92.236401} which relied on finite-size scaling of systems up to 1000 sites; for U = 4, $V_{BKT}$ = 1.877 by DMRG, and  $V_{BKT}$ = 1.89(1) by QMC, respectively.

The most obvious way to identify the central charge, and hence the peak, is with a regular curve fit; we also apply scaling to the middle for further gains in precision. The values of $c(V)_{max}$ extracted this way are shown in table \ref{ScaleToMiddleBKT}. One advantage of this approach is that all of the curve-fit parameters can be tabulated, including the Luttinger exponent K and the constant term in the entropy. As a result, as shown in table \ref{ScaleToMiddleBKT}, we found that the constant term in the entropy, $S_0$, is size-independent \footnote{Nishimoto \cite{Nishimoto2011} did not study the constant term at BKT transitions because his log-derivative approach removes it.}. The disadvantage of this approach is that about 32 sites are required to use scaling to the middle. This implies, for instance, that if the cold atoms under study are in a symmetric confining potential (for instance $U(x) \approx U + \Delta U x^2$), then nearly 32 atoms are needed to use scaling to the middle.

The second way of extracting $c(V)_{max}$ that we tested, is adapted from Ref.~\onlinecite{Nishimoto2011}, with results presented in table \ref{EntropyDerivative}. These results agree very well with the regular curve-fit method reported in table \ref{ScaleToMiddleBKT} and described above. It also agrees with our observed DMRG parameters (not tabulated here) in that at the computed BKT critical point, the entanglement (or equivalently, truncation error) is at a maximum at the BKT point. This method has the advantage, over curve-fitting, that fewer sites are needed to extract the critical point, providing easy access for experiments.

\begin{table}[ht]
\caption{The BKT point, determined by finding a maximum in central charge $c$ as a function of V, which was computed with the modified logarithmic derivative method. We noted an asymptotic \textit{increase} in central charge as truncation error was reduced, which may explain the inconsistency between the 128 site and 256 site results. The BKT transition is very sensitive to convergence.}
	\begin{ruledtabular}
	\begin{tabular}{ccc}
	
		\textbf{N} & \textbf{$V_{BKT}^{mid}$} & \textbf{$c_{BKT}^{mid}$}\\ \hline
		16         &     1.29(2)      &  0.89(1) \\ 
		32         &     1.57(1)      & 0.975(5) \\ 
		64         &     1.83(2)      &  1.06(1) \\ 
		128        &     1.96(2)      & 1.058(2) \\ 
		256        &      1.92(2)     &  1.027(2)\\ 
	\end{tabular}
\end{ruledtabular}
	\label{EntropyDerivative}
\end{table}

\section{Conclusions}

We have successfully demonstrated the precise identification of quantum critical points for the Extended Hubbard Model in 1D for both the second-order and the Berezinskii-Kosterlitz-Thouless transitions using nothing but the ground state von-Neumann entanglement entropy and results from Conformal Field Theory. Along the way we have introduced two refined methods for resolving quantum phase diagrams: ``Scaling to the Middle'' which provides improved measurement accuracy of any spatial curve fit on open boundary data, and an extended log-derivative approach for the study of central charge from open boundary data. Since the central charge exhibits a finite-size-effect peak at BKT transitions, it can then be used to identify such transitions from experimentally realistic system sizes. In combination with a CFT-fitted variance minimum, these tools enable reliable small-scale studies of numerical and experimental (i.e. cold-atom) entropy data.

In addition, we have identified the role played by soliton physics at the Gaussian critical point in the Extended Hubbard Model at half-filling; namely, it leads to an additional bosonic degree of freedom that appears as an addition to the central charge for systems up to about 100 sites in length. Lastly, we have observed the appearance of entanglement entropy oscillations for open boundaries that contradict previously observed effects in Luttinger Liquids.

\section{Acknowledgements}
We thank John Cardy for helpful contributions to our studies of the BKT transition (also see forthcoming publication) and Miles Stoudenmire for extensive discussions on ITensor. This research was supported in part by the NSF under grant DMR-1411345 and by UCR's GRMP fellowship (Winter 2016). This work used the Extreme Science and Engineering Discovery Environment (XSEDE) COMET at the San Diego Supercomputer Center through allocation TG-DMR170082 \cite{10.1109/MCSE.2014.80}.

%

\end{document}